\documentclass[namedreferences,natbib]{solarphysics}
\usepackage[optionalrh,solaromanenum]{spr-sola-addons} 
\usepackage{graphicx}                    
\usepackage{amssymb}                    
\usepackage{color}                       
\usepackage{url}                         
\usepackage{enumerate}

\begin{document}

\begin{article}

\begin{opening}

\title{Coronal Magnetic Field Evolution from 1996 to 2012: Continuous Non-Potential Simulations}

%
\author{A.R.~\surname{Yeates}
       }

%
\runningauthor{A.R. Yeates}
\runningtitle{Coronal Magnetic Field Evolution from 1996 to 2012}

%
  \institute{Department of Mathematical Sciences, Durham University, Durham, DH1 3LE, UK
                     email: \url{anthony.yeates@durham.ac.uk} 
             }

\begin{abstract}
Coupled flux transport and magneto--frictional simulations are extended to simulate the continuous magnetic field evolution in the global solar corona for over 15 years, from the start of Solar Cycle 23 in 1996.
By simplifying the dynamics, our model follows the build-up and transport of electric currents and free magnetic energy in the corona, offering an insight into the magnetic structure and topology that extrapolation-based models can not. 
To enable these extended simulations, we have implemented a more efficient numerical grid, and have carefully calibrated the surface flux transport model to reproduce the observed large-scale photospheric radial magnetic field, using emerging active regions determined from observed line-of-sight magnetograms. This calibration is described in some detail. In agreement with previous authors, we find that the standard flux transport model is insufficient to simultaneously reproduce the observed polar fields and butterfly diagram during Cycle 23, and that additional effects must be added. For the best--fit model, we use automated techniques to detect the latitude--time profile of flux ropes and their ejections over the full solar cycle. Overall, flux ropes are more prevalent outside of active latitudes but those at active latitudes are more frequently ejected. Future possibilities for space--weather prediction with this approach are briefly assessed.
\end{abstract}

%

\end{opening}

\section{Introduction}\label{sec:intro}

Modelling the Sun's coronal magnetic field over the full solar cycle is important because it acts as a driver of space--weather events, and changes significantly over the cycle, as well as from one cycle to the next.
It is fundamentally time--dependent, as manifested in the variations of almost all observed properties of the Sun, including sunspot number, magnetic flux, rates of flares and coronal mass ejections (CMEs), and even the total solar irradiance \citep{willson1991}.
The coronal magnetic field is driven by activity in the solar interior, including large-scale flows and convection in the photosphere as well as the periodic emergence of new active regions.

Previous studies of the coronal magnetic evolution over months to years have mostly used potential--field extrapolations \citep{altschuler1969,schatten1969}. From the space--weather viewpoint, these models give a first approximation of the Sun's open flux \citep{wang2002b} and the heliospheric current sheet \citep{hoeksema1983}, and have been coupled to models of the heliosphere \citep{arge2000,luhmann2002,pizzo2011}. They have also been applied to predict CME rates using the topology of magnetic nulls and ``breakout'' configurations \citep{cook2009}. However, their value for predicting flares and CMEs is limited since they allow no free energy, and the lack of electric currents leads them to underestimate the open flux, particularly during active periods \citep{riley2007,yeates2010c}.

Attempts to move beyond potential--field models and extrapolate more general magnetic equilibria suffer from the problem of non-uniqueness of magnetic topology for a given distribution of observed magnetic field in the photosphere. For example, the current--sheet source--surface (CSSS) model can better match the observed open flux by allowing certain forms of electric currents \citep{zhao1995,jiang2010b}, but these are chosen for mathematical convenience rather than any particular physical basis. On the other hand, when trying to extrapolate nonlinear force-free fields from photospheric vector magnetograms, different numerical extrapolation codes tend to produce different results \citep{derosa2009}. These difficulties have led to a more pragmatic approach for nonlinear force-free modelling of local structures such as coronal cavities or sigmoids, whereby the computation is initialised with a flux rope structure in the corona \citep{vanballegooijen2004,su2009,savcheva2012}.

In recent years, global magnetohydrodynamic (MHD) models that more realistically account for thermodynamic properties of the plasma have become practical, allowing for comparison with observed emission at various wavelengths \citep{lionello2009,rusin2010,downs2010}. However, it is not yet practical to simulate the temporal evolution of the corona for many months, and MHD models are generally limited to finding individual equilibria by relaxing from an initial condition appropriate for a given day. So far, the initial conditions for the magnetic field have been potential--field extrapolations. Thus the MHD models inherit the topology of the potential field, and do not account for the gradual build up of the magnetic topology over time, or any long-term memory of previous interactions that remains imprinted in the coronal field.

The approach of our model is to gain new insight into the magnetic structure and topology by not just extrapolating from photospheric data at a single time, but simulating the coronal field in a time-dependent way, using a simplified approximation to the real coronal evolution. The technique was originally introduced to study the formation of filaments \citep{vanballegooijen2000,mackay2000,mackay2001,mackay2005,mackay2006}, whose magnetic structure is non-potential and conjectured to depend on the build-up and transport of coronal magnetic helicity over time \citep{vanballegooijen1989}. It was extended to model the global corona by \citet{yeates2008}. The model effectively produces a continuous sequence of nonlinear force-free fields, in response to flux emergence and shearing by photospheric footpoint motions. As such, a particular magnetic topology is chosen automatically at each time step by the evolutionary history, thus providing a physically motivated solution to the non-uniqueness problem. Moreover, the model -- henceforth the NP (``non-potential'') model -- offers interesting possibilities for modelling and predicting space weather, because it allows for the build up and transport of free magnetic energy, electric currents, and magnetic helicity. In the model -- as is hypothesized in the real corona -- helicity tends to concentrate in flux rope structures overlying photospheric polarity--inversion lines. When too much helicity accumulates, the flux ropes ``erupt'' and are ejected through the outer boundary of the simulation domain \citep{mackay2006,yeates2009}.

The developments in this article are threefold. Firstly, we extend the simulations to a continuous 15 year evolution (Section \ref{sec:model}). Our previous study of solar cycle variations in the NP model was limited to six separate six--month runs \citep{yeates2010}. The new simulation allows for longer-term magnetic memory to take effect; our initial finding that this affects the chirality of high-latitude filaments has been described elsewhere \citep{yeates2012}. Secondly, we describe how the surface flux transport component of the model -- the lower boundary condition -- must be carefully calibrated to observations for such long simulations (Section \ref{sec:cal}). Thirdly, we consider the distribution of flux ropes and flux rope ejections over the 15-year simulation (Section \ref{sec:ropes}). We conclude in Section \ref{sec:outlook} with a brief discussion of the future prospects for space weather forecasting using this approach.

\section{Coronal Magnetic Model}\label{sec:model}

The non--potential (NP) model couples surface flux transport to magneto--frictional relaxation in the overlying corona \citep{vanballegooijen2000}.

\subsection{Formulation}\label{sec:form}
The large-scale mean coronal magnetic field $[{\bf B}_0=\nabla\times{\bf A}_0]$ is evolved by the induction equation
\begin{equation}
\frac{\partial{\bf A}_0}{\partial t} = {\bf v}_0\times{\bf B}_0 - {\bf E}_0,
\end{equation}
where we neglect ohmic diffusion and the mean electromotive force $[{\bf E}_0]$ describes the effect of unresolved small-scale fluctuations.
Following \citet{vanballegooijen2008}, we apply a hyperdiffusion
\begin{equation}
{\bf E}_0=-\frac{{\bf B}_0}{B_0^2}\nabla\cdot\big(\eta_4 B_0^2\nabla\alpha_0\big),
\end{equation}
where
\begin{equation}
\alpha_0 = \frac{{\bf B}_0\cdot{\bf j}_0}{B_0^2}
\end{equation}
is the current helicity density, with ${\bf j}_0=\nabla\times{\bf B}_0$ the current density, and $\eta_4=10^{11}\,{\rm km}^4{\rm s}^{-1}$. This form of hyperdiffusion preserves magnetic--helicity density $[{\bf A}_0\cdot{\bf B}_0]$ in the volume and describes the tendency of the magnetic field to relax to a state of constant $\alpha_0$ \citep{boozer1986,bhattacharjee1986}, although such a state is never reached in the global simulation.
The velocity is determined by the magneto--frictional technique \citep{yang1986,craig1986} as
\begin{equation}
{\bf v}_0 = \frac{1}{\nu}\frac{{\bf j}_0\times{\bf B}_0}{B_0^2} + v_{\rm out}(r){\bf e}_r.
\end{equation}
This replaces the full momentum equation and allows numerical solution of the model over months and years.
The first term enforces relaxation towards a force-free equilibrium $[{\bf j}_0\times{\bf B}_0=0]$. The second term is a radial outflow imposed only near the outer boundary $[r=2.5{\rm R}_\odot]$ to represent the effect of the solar wind radially distending magnetic field lines \citep{mackay2006}.

\subsection{Photospheric Boundary Condition}

On the photospheric boundary $r={\rm R}_\odot$, the magneto--frictional velocity is not applied, and instead the radial magnetic field $[B_{0r}]$ is evolved by the surface flux transport model \citep{sheeley2005,mackay2012}. In spherical polar coordinates $(r,\theta,\phi)$, the vector potential evolves according to
\begin{eqnarray}
\frac{\partial A_{0\theta}}{\partial t} &=& u_\phi B_{0r} - \frac{D}{{\rm R}_\odot\sin\theta}\frac{\partial B_{0r}}{\partial\phi} + S_\theta(\theta,\phi,t), \label{eqn:ath}\\
\frac{\partial A_{0\phi}}{\partial t} &=& -u_\theta B_{0r} + \frac{D}{{\rm R}_\odot}\frac{\partial B_{0r}}{\partial\theta} + S_\phi(\theta,\phi,t). \label{eqn:aph}
\end{eqnarray}
Here $D$ is a (constant) diffusivity modelling the random walk of magnetic flux owing to the changing supergranular convection pattern \citep{leighton1964}. In Section \ref{sec:cal}, we experiment with an additional exponential decay of $B_{0r}$ \citep{schrijver2002}. The differential rotation velocity $u_\phi=\Omega(\phi){\rm R}_\odot\sin\theta$ uses the observationally-determined \citet{snodgrass1983} profile
\begin{equation}
\Omega(\theta) = 0.18 - 2.3\cos^2\theta - 1.62\cos^4\theta\,{\rm deg}\,{\rm day}^{-1},
\end{equation}
written in the Carrington frame. For the basic meridional flow we assume the form of \citet{schuessler2006}, namely
\begin{equation}
u_\theta(\theta) = u_0\frac{16}{110}\sin(2\lambda)\exp\big(\pi - 2|\lambda | \big),
\end{equation}
where $\lambda=\pi/2 - \theta$ is latitude and $u_0$ is a constant controlling the flow amplitude. In Section \ref{sec:cal}, we experiment with activity-dependent perturbations of this flow profile.

The source terms $S_\theta$ and $S_\phi$ in Equations (\ref{eqn:ath}) and (\ref{eqn:aph}) represent the emergence of new active regions, and are necessary to maintain an accurate description of the observed surface $B_{0r}$ over the continuous 15 year simulation. Rather than specify a functional form for these source terms, we insert individual bipolar magnetic regions with properties chosen to match those in observed synoptic magnetograms \citep{yeates2007}. The inserted bipolar regions take an idealised three-dimensional form \citep{yeates2008}. In the future, we hope to incorporate more detailed models of the structure of individual active regions, built up in a time-dependent manner. For the simulations described here, we use synoptic normal--component magnetograms from US National Solar Observatory, Kitt Peak. Until 2003, these were taken with the older Vacuum Telescope, and from 2003 onward with Synoptic Optical Long-term Investigations of the Sun (SOLIS). We insert a total of 2040 bipoles between Carrington Rotation CR1911 (June 1996) and CR2122 (April 2012). We do not insert any bipoles to replace those missed during the three data gaps (CR2015\,--\,16, CR2040\,--\,41, CR2091).

\begin{figure} 
\centerline{\includegraphics[width=\textwidth]{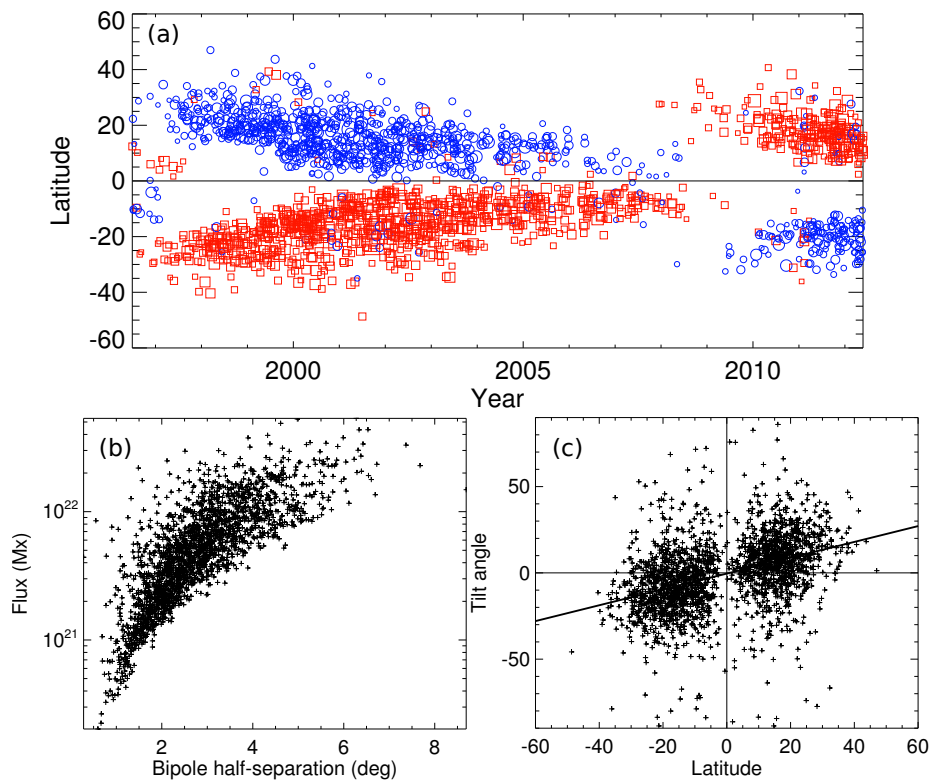}}
\caption{Properties of the 2040 magnetic bipoles determined from NSO/KP synoptic magnetograms: (a) locations in time and latitude, with colour/symbol showing polarity and symbol size proportional to flux; (b) bipole flux against bipole size; (c) tilt angles against latitude. In (c), the solid line is a linear fit to the measured tilt angles.}
\label{fig:bipoles}
\end{figure}

The 2040 bipolar regions are summarised in Figure \ref{fig:bipoles}. Figure \ref{fig:bipoles}(a) shows their latitude--time distribution, along with their leading/following polarity. The majority polarity reverses with each 11-year cycle, as it should. Figure \ref{fig:bipoles}(b) shows the range of sizes and fluxes of bipoles in our dataset. Our semi-automated technique identifies bipoles using only flux above $50\,{\rm gauss}$, thus imposing a lower size cut-off. Figure \ref{fig:bipoles}(c) shows the latitude distribution of bipole tilt angles $[\gamma]$ (the angle with respect to the Equator of the line connecting leading and following polarity centroids). The solid line shows a linear fit $\gamma=-0.41^\circ + 0.459\lambda$, which is consistent with magnetogram observations of \citet{wang1989}, who used NSO/KP data for Cycle 21, or \citet{stenflo2012}, who used SOHO/MDI magnetograms. White--light studies of tilt angles return a rather lower slope \citep{dasiespuig2010}. We feel that the distribution derived from magnetograms should be more appropriate for our application, although we find in Section \ref{sec:cal} that reducing the tilt angles by 20\,\% is a straightforward way to calibrate the flux--transport simulation without the need for more complex effects \citep{cameron2010a}. For the 20\,\% reduced tilt angles, the linear fit is $\gamma=-0.33^\circ + 0.367\lambda$.

\subsection{Numerical Methods}\label{sec:num}

The equations are solved for the vector potential $[{\bf A}_0]$ on a staggered grid, using a flux--conserving (finite--volume) scheme for the advection terms. To avoid the problem of grid convergence near the poles, we have newly incorporated a variable grid (see Appendix \ref{app:grid} for details). This greatly reduces the computational time. For the simulations described in this article, we use a resolution of 192 cells in longitude at the Equator, corresponding to an angular resolution of $1.875^\circ$, and $28$ cells in radius. At the polar grid boundary ($\theta\approx 0.67^\circ$), there are only 12 cells in longitude, owing to the variable grid. Boundary conditions are described in Appendix \ref{app:bc}. The coronal magnetic field is initialised using a potential--field extrapolation for 15 June 1996, derived from the synoptic magnetogram for CR1910 \citep{yeates2007}. The code is parallelised with OpenMPI and the full 5822-day run took approximately five days with 48 cores.

\section{Observational Calibration}\label{sec:cal}

Since we do not reset the photospheric magnetic field to observed magnetograms once the simulation has begun, it is important to calibrate the surface flux--transport model to reproduce the observed long--term evolution on the Sun. This is particularly true for simulations as long as 15 years, so we look at this issue in detail here. Unfortunately, it is a delicate balance between the properties of newly-emerging active regions and the chosen profiles for the transport processes of meridional flow and supergranular diffusion. Differential rotation is better constrained by observations and we do not consider its variation. Previous parameter studies of the flux--transport model have been undertaken by \citet{baumann2004}, and for Cycle 23 by \citet{schrijver2008} and \citet{jiang2011}, although here we are additionally constrained by the individual measured properties of the bipolar regions. Since there is no feedback of the coronal (magneto-frictional) component of the model on the surface flux--transport component, the calibration can be done by running only the latter.

To calibrate the surface simulations we use several numerical indicators for the photospheric magnetic field:
\begin{enumerate}[(i)]
\item Cross-correlation of total (unsigned) magnetic flux $[\Phi_\odot]$ with the observed time series.
\item Least value of $\Phi_\odot$ around the Cycle 23 Minimum (ca. 2009). 
\item Cross-correlation of axial dipole strength $B_{\rm ax}\equiv b_{1,0}$ with the observed time series.
\item Time of sign-reversal in the axial dipole strength.
\item Peak negative value of the axial dipole strength during Cycle 23.
\item Cross-correlation of equatorial dipole strength $B_{\rm eq}\equiv\sqrt{|b_{1,1}|^2 + |b_{1,-1}|^2}\equiv \sqrt{2}|b_{1,1}|$ with the observed time series.
\item Cross-correlation of longitude-averaged field $\langle B_{0r}\rangle\equiv\int_0^{2\pi}B_{0r}{\rm R}_\odot\sin\theta\,\mathrm{d}\phi$ with the observed butterfly diagram.
\end{enumerate}
Here $b_{l,m}$ are the complex--valued spherical--harmonic components of $B_{0r}$. Because of the smoothing effect of supergranular diffusion, the total simulated flux $[B_{0r}]$ tends to be lower than the total magnetogram flux. To account for this, our cross-correlations (a) and (g) use a smoothed version of the observed magnetogram, obtained by removing spherical harmonic components above order $\ell=128$. Higher--resolution simulations in future will need to incorporate a more realistic random walk model for discrete flux concentrations \citep[{\it e.g.},][]{worden2000,schrijver2001}.
To facilitate cross-correlation of time series in (a), (c), and (f), the simulation data are first interpolated to the same times as the magnetogram observations. For cross-correlation of the two-dimensional butterfly diagrams, the simulation data are first interpolated to the same times and latitudes as the observations.

\begin{figure} 
\centerline{\includegraphics[width=\textwidth]{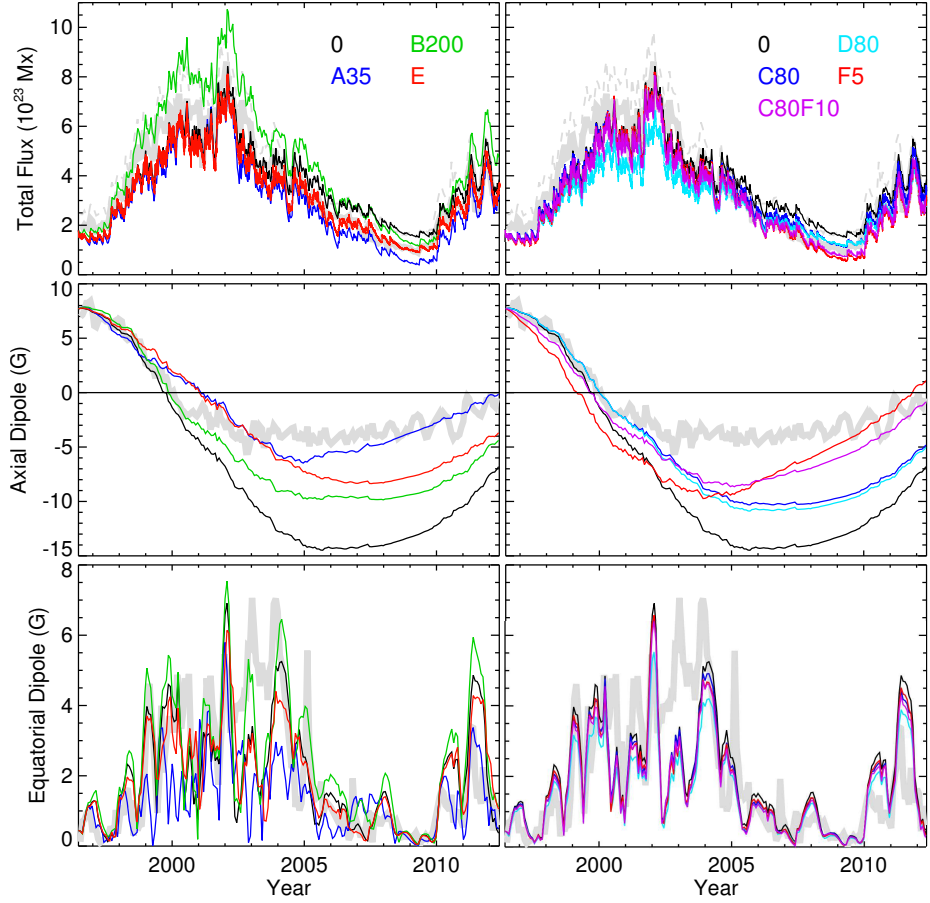}}
\caption{Time series of total unsigned photospheric flux $[\Phi_\odot]$, axial dipole component $[B_{\rm ax}]$, and equatorial dipole component $[B_{\rm eq}]$ for the various surface calibration runs. Observed (NSO/KP) values are shown by thick grey lines (in the top row, the dashed grey line shows value from the original magnetograms and the solid grey line that from smoothed magnetograms).}
\label{fig:series}
\end{figure}

\begin{figure} 
\centerline{\includegraphics[width=\textwidth,clip=]{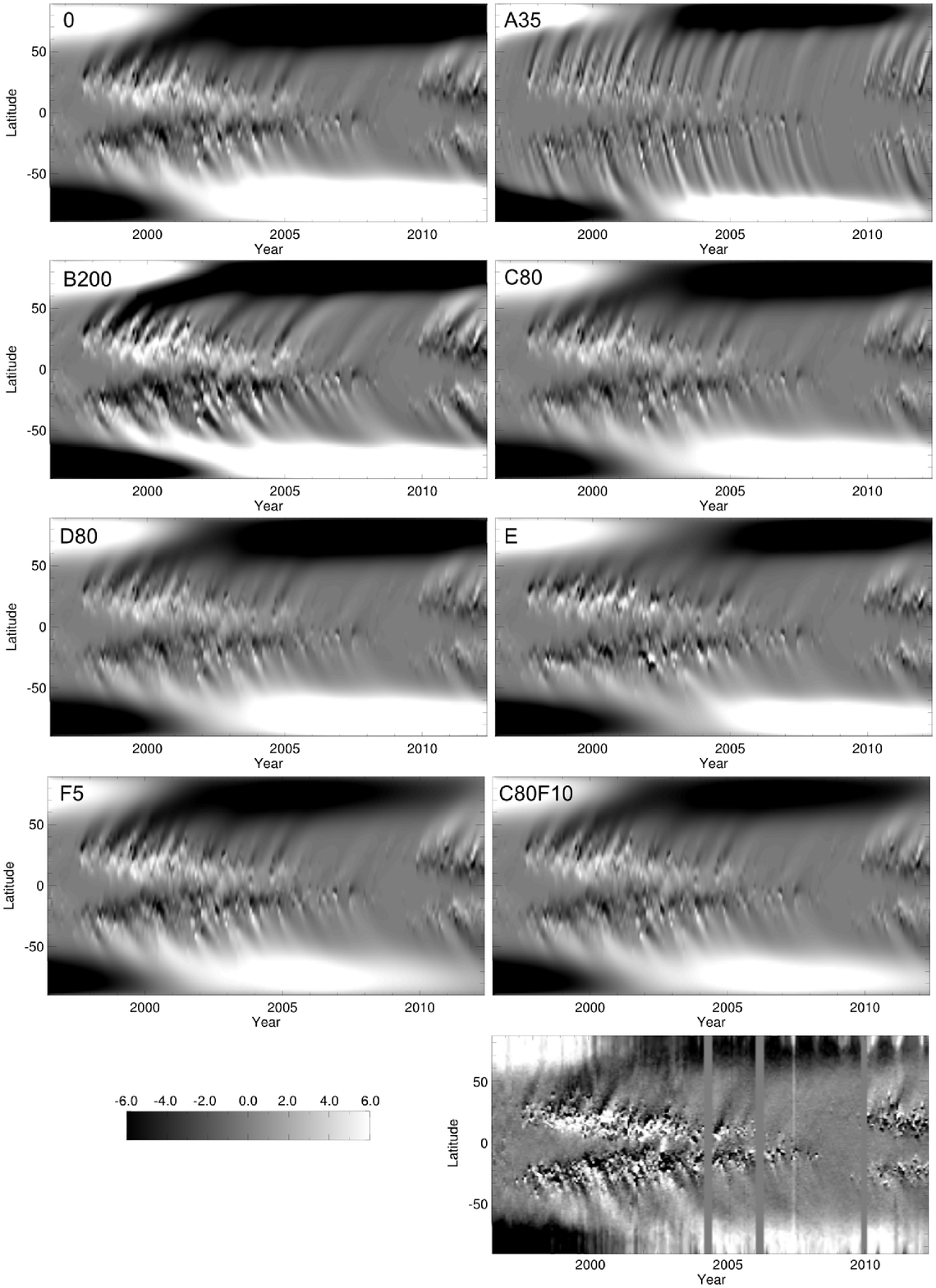}}
\caption{Magnetic butterfly diagrams showing the longitude-average $\langle B_{0r}\rangle$ for each calibration run, and for the observed KP magnetograms (bottom right). In each case the greyscale saturates at $\pm 6\,{\rm G}$. Data gaps and problems with high-latitude measurements are evident in the observed butterfly diagram.}
\label{fig:bfly}
\end{figure}

Table \ref{tab:cors} lists values of the numerical indicators for a representative set of test runs, which are described in more detail below. For comparison, the total unsigned flux $[\Phi_\odot]$, axial dipole strength $[B_{\rm ax}]$, and equatorial dipole strength $[B_{\rm eq}]$ for each run are shown in Figure \ref{fig:series}. Butterfly diagrams of $\langle B_{0r}\rangle$ in each run are shown in Figure \ref{fig:bfly}.

\begin{table}
\caption{Performance of surface flux transport calibration runs.}
\label{tab:cors}
\begin{tabular}{rrrrrrrr}     
\hline
Run	& \multicolumn{7}{c}{Indicators}\\
    & (i) & (ii) $[10^{23}\,{\rm Mx}]$ & (iii) & (iv) [years] & (v) [gauss] & (vi) & (vii)\\
\hline
Observed &---  &1.520&---  &2000.04&-5.47 &---  &--- \\
Smoothed &---  &0.895&---  &---    &---   &---  &0.945\\
\hline
0        &0.941&1.314&0.934&1999.71&-14.49&0.688&0.693  \\
A35      &0.965&0.392&0.938&2001.19&-6.47 &0.570&0.590\\
B200     &0.968&1.146&0.951&1999.93&-9.85 &0.684&0.709\\
C80      &0.960&1.119&0.933&2000.08&-10.32&0.709&0.711\\
D80      &0.954&1.147&0.931&2000.08&-10.88&0.687&0.708\\
E        &0.967&0.898&0.898&2001.04&-8.36 &0.673&0.695\\
F5       &0.966&0.553&0.913&1999.19&-9.74 &0.682&0.706 \\
C80F10   &0.967&0.702&0.966&1999.64&-8.65 &0.706&0.726 \\
\hline
\end{tabular}
\end{table}

\subsection{Reference Case}

Consider first the reference case Run 0, with peak meridional flow $u_0=11\,{\rm m}\,{\rm s}^{-1}$ \citep[consistent with Cycle 23 observations by][]{hathaway2010} and diffusivity $D=450\,{\rm km}^2{\rm s}^{-1}$. The time series for Run 0 are shown by black curves in all panels of Figure \ref{fig:series}, while corresponding values for the observed magnetograms are shown by thick grey lines. It is clear that $\Phi_\odot$ is too high from about 2004 onwards, during the declining phase. This is clearly caused by an excess in $B_{\rm ax}$ rather than $B_{\rm eq}$. Indeed, the butterfly diagram (Figure \ref{fig:bfly}) shows that Run 0 builds up unrealistically large polar caps. The low polar field of Cycle 23 is not reproduced by this standard model. A second inconsistency in Run 0 is a deficit in the equatorial dipole strength $B_{\rm eq}$ around 2002\,--\,2004. This is present in all of the calibration runs and appears to be due to an underestimate of emerged flux in a handful of particular active regions near the Equator, to which $B_{\rm eq}$ is sensitive. This is likely caused by our simplified method of extracting bipolar regions, but fortunately does not have a significant effect on $\Phi_\odot$ or on the later polar field produced.

\subsection{Meridional Flow Speed}

A possible solution to the excess polar field problem is to retain the standard flux transport model but to alter the meridional flow speed $[u_0]$ or diffusivity $[D]$. Differential rotation is better constrained and does not significantly affect the dipole strengths \citep{baumann2004}. In order to reduce the polar field production, one must increase $u_0$, so as to reduce cancellation at the Equator \citep{devore1984}. In Run A35, the required speed of $u_0=35\,{\rm m}\,{\rm s}^{-1}$ is used to bring $B_{\rm ax}$ down to the observed level (Figure \ref{fig:series}). But in addition to being inconsistent with observations \citep{hathaway2010}, this causes a very late reversal of $B_{\rm ax}$, and too low a level of $B_{\rm eq}$ over much of the cycle. The resulting butterfly diagram is also poorly correlated with observations (Figure \ref{fig:bfly}).

\citet{schrijver2008} and \citet{jiang2011} were able to improve the match of their simulations to observed dipole moments/polar fields by changing the meridional flow. However, when we run a simulation with our measured bipole properties and the diffusivity ($250\,{\rm km}^2{\rm s}^{-1}$) and meridional flow profile of \citet{jiang2011}, we find that the axial dipole strength is close to that of Run 0, even with their 55\,\% enhanced flow speed ($17\,{\rm m}\,{\rm s}^{-1}$). Moreover, correlation with the observed butterfly diagram is again poor. If we implement the meridional--flow profile of \citet{schrijver2008}, with a stronger latitudinal gradient in $u_\theta$ at the Equator, and $u_0=15\,{\rm m}\,{\rm s}^{-1}$, we still obtain similar results. Note that neither of those studies used observations of individual bipolar magnetic regions, as we do here.

\subsection{Supergranular Diffusivity}

In Run B200, we instead reduce $D$ to $200\,{\rm km}^2{\rm s}^{-1}$. This produces slightly high $\Phi_\odot$ compared with the observations throughout the cycle, although the butterfly diagram and $B_{\rm eq}$ are a better fit to the observations. In the declining phase, $B_{\rm ax}$ is still rather high. Here we note a discrepancy with the parameter study by \citet{baumann2004} for our $u_\theta$ profile and bipole properties: as we increase $D$, the peak axial dipole $B_{\rm ax}$ \emph{increases}, whereas they found the maximum polar field strength to \emph{decrease} over the same range of $D$.

\subsection{Bipole Properties}

Although our bipolar regions are constrained individually by observations, we also consider systematically modifying their properties. One can reduce the polar field either by reducing the bipole fluxes, or by reducing the average tilt angle so that each region contributes less to the axial dipole \citep{wang2000}. In Run C80, all tilt angles were reduced by $20\,\%$, while in Run D80 all bipole fluxes were reduced by $20\,\%$. Both runs produce a comparable improvement in $B_{\rm ax}$, with accurate reversal times. Overall, Run C80 produces better results than Run D80, because the latter causes too much reduction in $B_{\rm eq}$ and consequently in $\Phi_\odot$, particularly during active periods. \citet{jiang2011} found that a 28\,\% decrease in their tilt angles (as compared to previous cycles) produced a reasonable polar field in Cycle 23. However, even in Run C80 the peak of $B_{\rm ax}$ remains too strong compared to the observations, so we are led to consider an additional change to the model.

\subsection{Exponential Flux Decay}

In run F5, we try adding additional exponential decay terms of the form $-A_{0\theta}/\tau$ and $-A_{0\phi}/\tau$ to Equations (\ref{eqn:ath}) and (\ref{eqn:aph}) respectively, leading to an exponential decay on a timescale $\tau=5\,{\rm years}$. Such a decay has previously been introduced by \citet{schrijver2002} in flux--transport simulations of the past 340 years, in order to maintain regular polar--field reversals when cycles vary in strength from one to the next. \citet{baumann2006} have introduced a similar enhanced decay, proposing the physical explanation to be volume diffusion of the surface magnetic field in the three-dimensional solar interior. Figure \ref{fig:series} shows that this decay is able to bring $B_{\rm ax}$ down to the observed level in 2010 while maintaining a good correlation with the observed butterfly diagram, although it has the side-effect of bringing the reversal time of $B_{\rm ax}$ too early. By combining a weaker exponential decay term ($\tau=10\,{\rm years}$) with 20\,\% reduced tilt angles (as in Run C80), Run C80F10 produces a better compromise and an even stronger correlation with the observed butterfly diagram. This is the run chosen for driving the coronal simulations in Section \ref{sec:ropes}, and used by \citet{yeates2012}.

\subsection{Time-varying Meridional Flow}

An alternative improvement to the flux--transport model is to introduce temporal and spatial fluctuations in the meridional flow. Either a high-latitude countercell \citep{jiang2009} or converging flows towards active regions \citep{jiang2010} can reduce the resulting polar field. Although we do not carry out an exhaustive investigation here, we illustrate (Run E) the effect of such converging flows using the method of \citet{cameron2010}. An axisymmetric perturbation is added to the steady flow $u_\theta$ to model the net effect of non-axisymmetric perturbations \citep{derosa2006}, and takes the form
\begin{equation}
u'(\lambda,t)=c_0\left(\frac{\mathrm{d}}{\mathrm{d}\lambda}\langle|B_{0r}|\rangle(\lambda,t)\right),
\end{equation}
where $\langle|B_{0r}|\rangle(\lambda,t)$ is the longitudinal average of $|B_{0r}|$, to which we also apply a Gaussian smoothing in latitude at each time. Following \citet{cameron2010} we set $c_0=10\,{\rm m}\,{\rm s}^{-1}{\rm gauss}^{-1}\,{\rm deg}$. The Gaussian smoothing is implemented (at each time step of the main simulation) by taking 80 one-dimensional (in $\lambda$) diffusive steps with diffusion coefficient $2.4\times 10^7\,{\rm km}^2\,{\rm s}^{-1}$. \citet{cameron2010} show that this form of $u'(\lambda,t)$ leads naturally to variation of the $P_1^2(\cos\theta)$ component of the total meridional flow over the cycle. It thus reproduces both the observations of \citet{hathaway2010}, who found a variation of the $P_1^2$ component using MDI tracking, and \citet{basu2010}, who both substantiated the Hathaway and Rightmire result and found evidence for inflows towards active regions, using MDI helioseismology. For our bipole properties, Figure \ref{fig:series} shows that these flow perturbations lead to reasonable $\Phi_\odot$, and reasonable $B_{\rm ax}$ later in the cycle. But the reversal of $B_{\rm ax}$ is much later than observed, because poleward transport is reduced during the rising phase of the cycle. This leads also to poorer correlation with the observed butterfly diagram. Hence we do not pursue this route here, although temporal variations in meridional flow will be important to develop in the future, and particularly to take into account variations between different cycles \citep{wang2002,schrijver2008}.

\section{Flux Ropes and Eruptions}\label{sec:ropes}

As an application of the NP model, we focus here on the formation and ejection of magnetic--flux ropes over the 15--year simulation, driven by the optimal flux transport Run C80F10. Twisted flux ropes form naturally in the simulation when surface motions concentrate magnetic helicity above polarity inversion lines in the photospheric field. Previously, we have developed automated techniques to detect them, and have analysed the effect of the various coronal simulation parameters on their formation and ejection \citep{yeates2009}. We have also undertaken a detailed comparison with CME source regions observed in the extreme ultraviolet \citep{yeates2010}. \citet{yeates2010b} looked at how the flux rope statistics differed in different phases of Cycle 23, using six-month ``snapshots''. Here, we present the distribution of flux ropes and flux--rope ejections over the continuous 15-year simulation. This is intended to be an illustration of the model capabilities, rather than a detailed parameter study. In particular, the model is not yet suitable for prediction of individual space--weather events (see Section \ref{sec:outlook}).


\subsection{Definition and Automated Identification}

\begin{figure} 
\centerline{\includegraphics[width=0.4\textwidth]{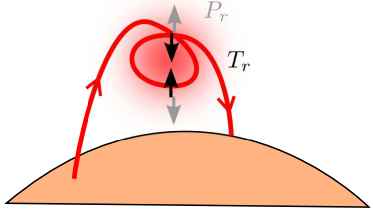}}
\caption{Flux rope points are identified by an inward magnetic tension force $[T_r]$ and an outward magnetic pressure force $[P_r$].}
\label{fig:ropeforce}
\end{figure}

\begin{figure} 
\centerline{\includegraphics[width=\textwidth]{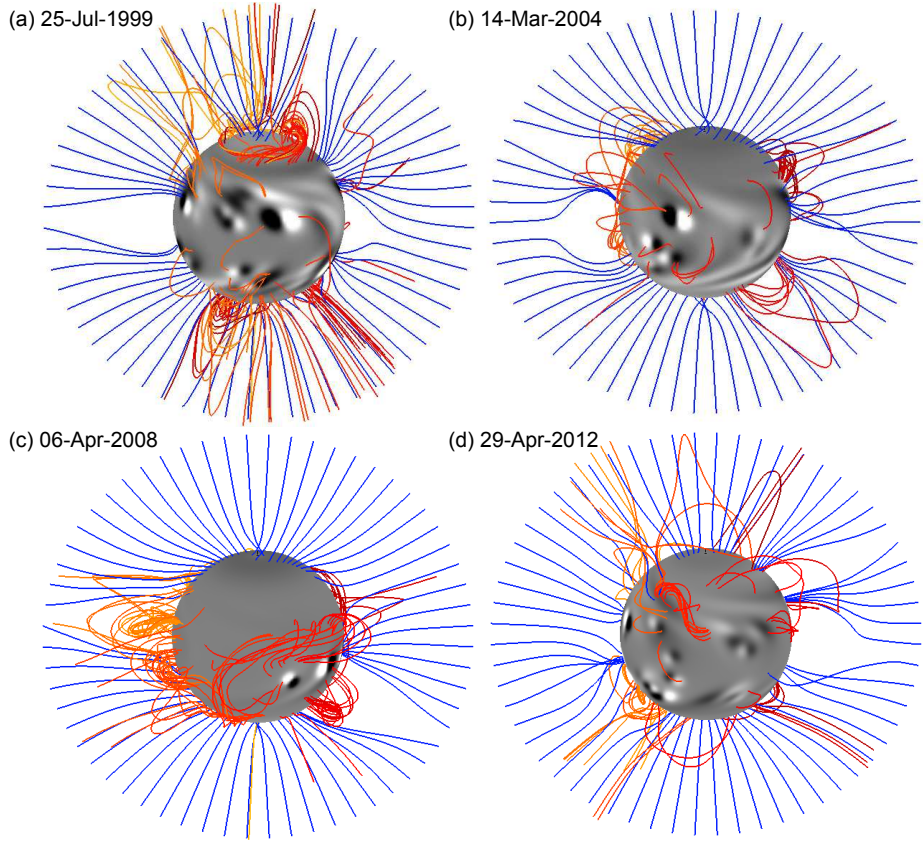}}
\caption{Snapshots of the NP model at a sequence of times during Run C80F10, including (a) Cycle 23 Maximum prior to polar reversal, (b) the declining phase, (c) Cycle 23 Minimum, and (d) Cycle 24. In each case, blue field lines are traced down from the source surface $r=2.5{\rm R}_\odot$, while red field lines are traced from (a subset of) flux rope points.}
\label{fig:snaps}
\end{figure}

Identifying magnetic--flux ropes in complex three-dimensional magnetic fields is a non-trivial task. We do this once per day during the course of the simulation, using a slight modification of the automated technique described by \citet{yeates2009}. The technique works by first identifying flux--rope \emph{points} on the numerical grid satisfying certain criteria, before grouping these points into flux \emph{ropes} using a clustering algorithm. We have settled on the following criteria for defining flux--rope points. At each point on the computational grid, we compute the normalised vertical magnetic tension force and pressure gradient,
\begin{equation}
T_r = \frac{{\rm R}_\odot}{B_0^2}{\bf B}_0\cdot\nabla\left(\frac{B_{0r}}{\mu_0}\right), \qquad P_r = -\frac{{\rm R}_\odot}{B_0^2}\frac{\partial}{\partial r}\left(\frac{B_0^2}{2\mu_0} \right).
\end{equation}
A grid point $(r_i,\theta_j,\phi_k)$ is then selected if it satisfies the following five conditions:
\begin{eqnarray}
P_r(r_{i-1},\theta_j,\phi_k) &<& -0.4,\\
P_r(r_{i+1},\theta_j,\phi_k) &>& 0.4,\\
T_r(r_{i-1},\theta_j,\phi_k) &>& 0.4,\\
T_r(r_{i+1},\theta_j,\phi_k) &<& -0.4,\\
|{\bf j}_0\cdot{\bf B}_0| &>& \alpha^*B_0^2,
\end{eqnarray}
where $\alpha^*=0.7\times 10^{-8}\,{\rm m}^{-1}$. Thus a flux rope is defined as a twisted structure where magnetic pressure acts outward from the flux rope axis and magnetic tension acts inward (Figure \ref{fig:ropeforce}). As an illustration, Figure \ref{fig:snaps} shows four snapshots of the corona in Run C80F10, with red/orange magnetic--field lines traced from identified flux--rope points.

To identify flux--rope ejections, we use the automated post-processing procedure of \citet{yeates2009}. This selects those flux--rope points with $v_{0r}>0.5\,{\rm km}\,{\rm s}^{-1}$ in the magneto-frictional code and clusters them both spatially and temporally into separate ejection events. To be classed as separate events, clusters must be separated by at least five days in time and have at least eight points (on the grid resolution used here). This yields a list of times, locations, and sizes of flux rope ejections during the simulation.

\subsection{Latitude--Time Distribution}

\begin{figure} 
\centerline{\includegraphics[width=\textwidth]{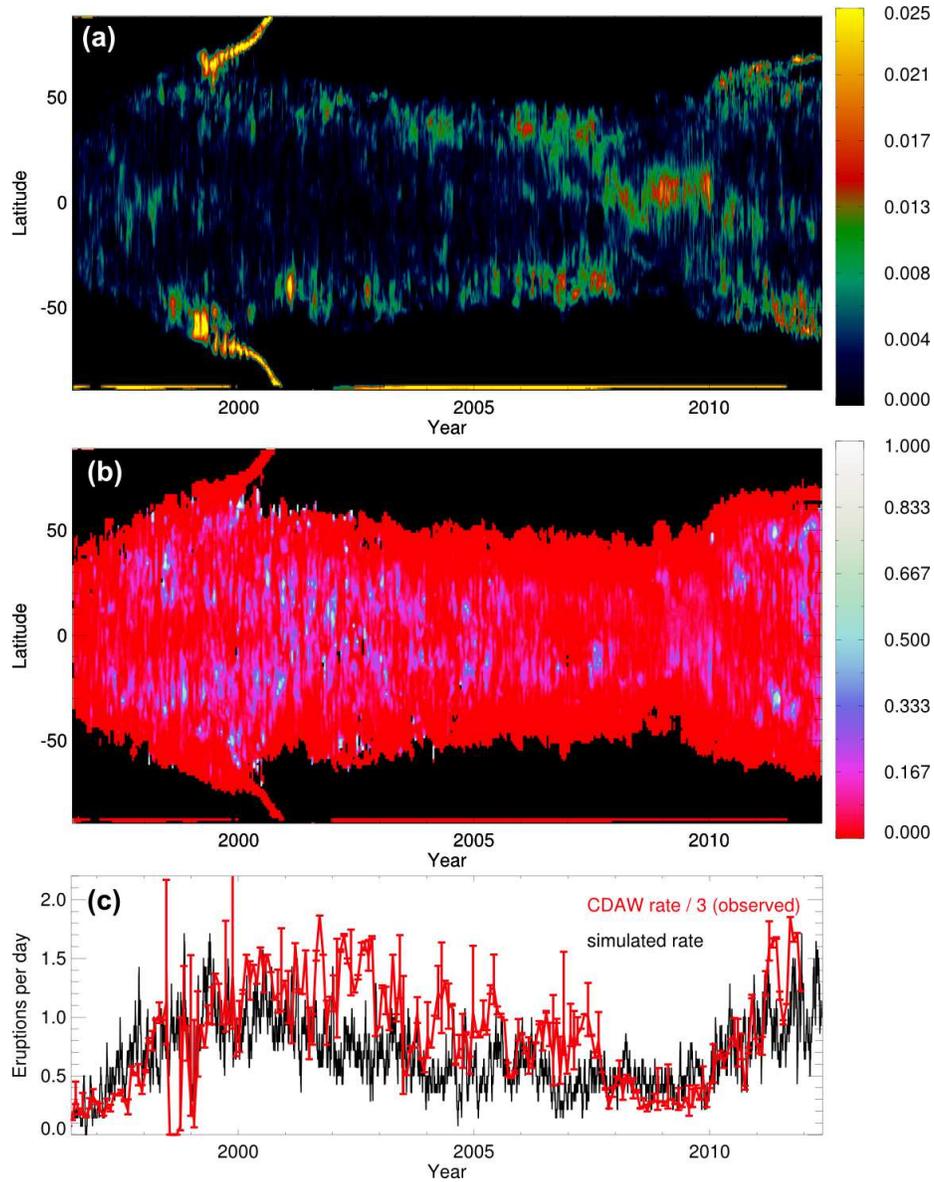}}
\caption{Latitude--time distribution of (a) flux ropes and (b) flux--rope eruptions, and (c) time series of the flux--rope eruption rate. In (a), the colour scale shows flux rope \emph{filling factor}, while in (b), the colour scale shows the fraction of flux rope points erupting in a given latitude--time bin. Bins with no flux rope points are coloured black. In (c), the observed CME rate from the CDAW catalogue \emph{divided by three} is plotted in red (see text).}
\label{fig:ropebfly}
\end{figure}

At any one time, there are multiple flux ropes, of varying sizes, present in the simulation. To show how these vary over latitude and time, Figure \ref{fig:ropebfly} summarises the results of the automated detection routines applied to Run C80F10.

Figure \ref{fig:ropebfly}(a) shows the distribution of flux ropes in time and latitude. The quantity plotted is the \emph{filling factor} of flux ropes, namely the proportion of grid points at each latitude and time that are identified as flux rope points. There are two features to notice about the flux--rope distribution. Firstly, the latitude range of flux ropes reflects the extent of PILs on the solar surface, and varies over the cycle. The distribution of PILs reaches its broadest extent during the ``rush-to-the-poles'' in the run up to polar field reversal in 2000\,--\,2001, after which high-latitude flux ropes disappear again. The second feature to notice is that the filling factor is greater outside of active regions, either at high latitudes or during the Minimum period from 2007\,--\,2010 when there were few active regions present. This anti-correlation with the magnetic butterfly diagram fits the picture of flux ropes forming as a result of gradual transport and build-up of helicity by surface motions. In fact, the number of individual flux ropes does increase by a factor of about 1.5 during Solar Maximum \citep{yeates2010b}, but this is because a greater number of smaller flux ropes are present in and around active regions. Larger ropes are found in decaying flux regions where helicity has had time to concentrate.

This observed distribution of flux ropes shares some similarities with observed butterfly diagrams of solar filaments, which also overlie PILs \citep{dazambuja1948,mouradian1994}. Namely, the overall latitudinal range and the increase in the number of filaments with solar activity. However, \citet{mouradian1994} find that, around the Minimum of Cycle 21, filaments persisted only at higher latitudes and not near the Equator. This appears to be at odds with the simulated distribution of flux ropes in Figure \ref{fig:ropebfly}(a) in the years 2008\,--\,2010, as seen in Figure \ref{fig:snaps}(c). This requires further investigation.

Figure \ref{fig:ropebfly}(b) shows the proportion of flux--rope points found to be erupting, within each latitude--time bin. Empty bins are black. Notice that the pattern is predominantly the inverse of Figure \ref{fig:ropebfly}(a), and now correlates with the magnetic butterfly diagram. So we find that, although less space is filled with flux ropes at active latitudes, the flux ropes that are present at active latitudes are more likely to erupt. Hence the eruption rate is modulated by solar activity (black curve in Figure \ref{fig:ropebfly}c). This agrees with the findings of \citet{yeates2010b}, who found that the flux--rope eruption rate increased by a factor of eight between 1996 and 1999, comparable to the relative increase of the bipole emergence rate, or of the total magnetic energy.

Note that the rate in 1996 may be underestimated -- both in Figure \ref{fig:ropebfly} and in the earlier simulation -- due to a lack of helicity stored in the initial configuration. \citet{yeates2012} show how the helicity on high-latitude PILs depends not only on \emph{in situ} generation by differential rotation, but also on the gradual poleward transport of helicity from lower latitudes. Thus there is a time delay for the build up of non-potential structure in the corona. Indeed the subsequent Cycle 23 Minimum (2007\,--\,2009) shows a higher floor in the eruption rate of roughly 0.4 per day. So if the simulation were initialised earlier, we might expect a higher eruption rate in 1996. This highlights the importance of long-term memory in the coronal magnetic topology. The gradual poleward transport of helicity also explains the relatively infrequent eruptions at higher latitudes. While such eruptions are present, for example around the ``detachment'' of the polar crowns in 2000, these structures tend to encircle much of the Sun. Once helicity is removed in an eruption, it takes time to build again.

\section{Outlook for Space Weather Prediction}\label{sec:outlook}

Flux--rope eruptions in the NP model can give us insight into the origin of CMEs,  although further development is needed before the model can enter the realm of operational space--weather prediction. In the first place, the peak eruption rate of 1.5 per day in the simulation is substantially lower than the rate of observed CMEs, by approximately a factor three when compared with the manually compiled Coordinated Data Analysis Workshops catalogue \citep[CDAW,][]{yashiro2004}, which is based on data from Large Angle and Spectrometric Coronagraph \citep[LASCO,][]{brueckner1995}. The CDAW rate, divided by three, is shown by the red curve in Figure \ref{fig:ropebfly}(c). We have filtered out CMEs with apparent width less than $15^\circ$ or greater than $270^\circ$, and determined error bars by taking into account data gaps \citep{stcyr2000,yeates2010b}.

The parameter study of \citet{yeates2009} shows that the eruption rate may be increased either by reducing the coronal diffusivity in the simulation (equivalent to $\eta_4$ here), or by increasing the hemispheric twist imbalance of the emerging bipolar regions. However, the latter is constrained by observations of filament chirality \citep{yeates2008}, and for reasonable values of these parameters the eruption rate remains too low.

Many missed eruptions are likely in active regions. In the present simulations, emerging regions are treated as idealised bipoles, and the effect of this simplification is likely to be most pronounced during the early stages of an active region's lifetime. Our model was designed to follow the long-term evolution of magnetic helicity, and cannot reproduce multiple eruptions in rapid succession from a single active region. Nor are the idealised bipoles a spatially accurate model for more complex $\delta$-spot regions. These limitations prevent us from making either spatial or temporal predictions of individual CMEs \citep{yeates2010}.

Nevertheless, more detailed modelling of active--region structure and emergence could in principle be incorporated into the NP model in future. The difficulty is that one needs a description of the three-dimensional structure of emerging regions. The solution may be either to impose time-dependent electric fields on the solar surface \citep{fan2007,fisher2010}, or to locally drive the simulation from higher-cadence magnetograms in place of the flux--transport model \citep{mackay2011,cheung2012}, although these techniques are still under development.

We believe that the time-dependent NP model is worth pursuing because the inherent magnetic memory in the corona implies a degree of predictability. While flares or CMEs from newly-emerged active regions are unlikely to be predicted more than a day or two before the active region appears on the surface, CMEs originating from decaying active regions offer scope for longer-term prediction using models such as this. Finally, it should be noted that the NP model includes only the magnetic field and not other plasma properties. Hence it can predict only the initiation of flux--rope ejections, not their subsequent dynamical evolution. For individual events, this evolution can be followed in MHD models \citep[e.g.,][]{manchester2004}.

%
 \appendix   

\section{Computational Grid} \label{app:grid}

\begin{figure} 
\begin{center}
\includegraphics[width=0.4\textwidth]{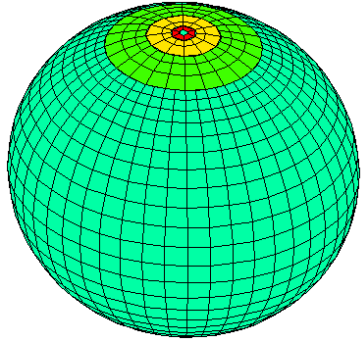}
\end{center}
\caption{Example of the variable grid with 48 cells at the equator (compared to 192 in the actual simulations) and seven sub-blocks (compared to nine).}
\label{fig:blocks}
\end{figure}

\begin{figure} 
\begin{center}
\includegraphics[width=0.65\textwidth]{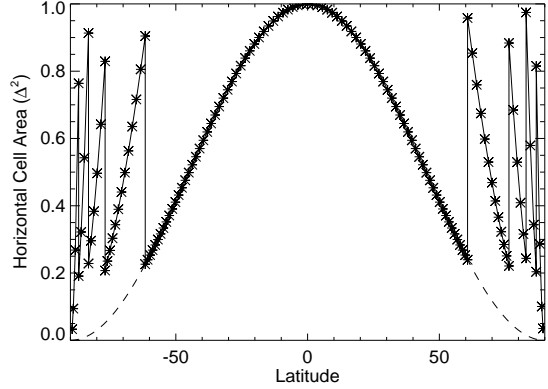}
\end{center}
\caption{Horizontal cell area (relative to that at the Equator) as a function of latitude, where symbols (joined by solid lines) show the variable grid and dashed lines a uniform grid with $\mathrm{d}x=\mathrm{d}y=1$ everywhere.}
\label{fig:grid}
\end{figure}

Our computational grid is divided into latitudinal sub-blocks, each of which has uniform spacing in the stretched variables
\begin{equation}
x=\phi/\Delta, \quad y=-\log(\tan(\theta/2))/\Delta, \quad z=\log(r/{\rm R}_{\odot})/\Delta
\end{equation}
\citep{vanballegooijen2000}, where $\Delta$ is the equatorial grid spacing in longitude $[\phi]$. The horizontal cell-sizes are $\mathrm{d}x=\mathrm{d}y=1$ for the equatorial sub-block, and double in each sub-block towards the poles (Figure \ref{fig:blocks}). The vertical cell size is $dz=1$ for all sub-blocks. This introduction of sub-blocks with different spacing counters the problem of grid convergence toward the poles, since the horizontal cell area is $\Delta^2r^2\sin^2\theta\,\mathrm{d}x\mathrm{d}y$. The sub-block boundaries in latitude are determined so that each grid cell is as large in horizontal area as possible, while never exceeding the equatorial cell area $\Delta^2r^2$ (Figure \ref{fig:grid}). With a longitudinal resolution of 192 cells at the Equator and 12 at the poles, there are nine sub-blocks. (The number 12 is chosen due to the parallel architecture used.) The total number of grid cells in $(x, y)$ is 18\,936, as compared to 63\,744 for a uniform, single-block grid with unit spacing.

A complication arising from the variable grid is that the different sub-blocks need to communicate ghost values of $B_{0r}$ and $B_{0\phi}$ with one another at each timestep. This is analogous to the inter-level communications in Adaptive Mesh Refinement (AMR) codes, except that our grid is fixed in time. Restriction (from fine to coarse sub-blocks) is the simpler process, for which we use an area-weighted average over fine grid cells \citep{balsara2001}. Prolongation (from coarse to fine sub-blocks) is trickier since the coarse--block solution must be interpolated to get the fine--block ghost values at intermediate locations. We follow the Taylor expansion method of \citet{vanderholst2007}, using monotonic van Leer slope estimates  \citep{evans1988}. These slopes are also computed throughout the grid and used for slope-limiting in the advection terms, in order to prevent spurious oscillations near sharp gradients. Our numerical tests indicate that numerical diffusion due to this ``upwinding'' is negligible compared to the physical diffusion $D$. Finally, after computing the update $\partial{\bf A}_0/\partial t$ within each sub-block, we replace boundary values on coarser sub-blocks with those derived from finer sub-blocks. This is analogous to the ``flux correction'' of \citet{berger1989}.

\section{Global Boundary Conditions} \label{app:bc}

The staggered grid requires ghost--cell values of two components of ${\bf B}_0$ outside each boundary. In longitude the domain is simply periodic.
On the photosphere $r={\rm R}_\odot$, we fix ghost cell values of $B_{0\theta}$, $B_{0\phi}$ by requiring that $v_{0r}=0$ on $r={\rm R
}_\odot$ in the magneto--frictional model. On the outer boundary $r=2.5{\rm R}_\odot$, we impose a radial outflow velocity \citep[see][]{yeates2010c} that models the effect of the solar wind radially extending field lines, while still allowing horizontal fields to escape during flux--rope ejections. The resulting ghost--cell values of $B_{0\theta}$, $B_{0\phi}$ do not play an important role and are simply set by zero-gradient conditions. On the latitudinal boundaries (located at approximately $\pm 89.33^\circ$ latitude) we need ghost cell values for $B_{0r}$ and $B_{0\phi}$. The latter are simply set to zero, while the ghost values of $B_{0r}$ are chosen to satisfy Stokes' Theorem given the integral of $A_{0\phi}$ around the latitudinal boundary. 

%
\begin{acks}
The author thanks NSO for support to attend the 26th Sac Peak workshop,  D.H.~Mackay for useful discussions during this research, and the anonymous referee for helpful suggestions. Numerical simulations used the STFC and SRIF funded UKMHD cluster at the University of St Andrews. Magnetogram data from NSO/Kitt Peak were produced cooperatively by NSF/NOAO, NASA/GSFC, and NOAA/SEL, and SOLIS data are produced cooperatively by NSF/NOAO and NASA/LWS. The observed CME catalogue used is generated and maintained at the CDAW Data Center by NASA and The Catholic University of America in cooperation with the NRL. SOHO is a project of international cooperation between ESA and NASA. 
\end{acks}

%
%
 \bibliographystyle{spr-mp-sola-cnd} 
 \bibliography{yeates}  
%
%
%
%

\end{article} 
\end{document}